\documentclass{aa}
\usepackage{natbib}
\bibpunct{(}{)}{;}{a}{}{,} 
\usepackage{graphicx}
\usepackage{txfonts}
\usepackage{hyperref}
\usepackage{upgreek}

\begin{document} 

   \title{LoTSS/HETDEX: Disentangling star formation and AGN activity in gravitationally-lensed radio-quiet quasars}
   \titlerunning{Gravitationally-lensed radio-quiet quasars in LoTSS/HETDEX}
   \author{
   H.~R.~Stacey\inst{1,2}\thanks{h.r.stacey@astro.rug.nl},
   J.~P.~McKean\inst{1,2},
   N.~J.~Jackson\inst{3},
   P.~N.~Best\inst{4},
   G.~Calistro~Rivera\inst{5},
   J.~R.~Callingham\inst{1},
   K.~J.~Duncan\inst{5},
   G.~G\"urkan\inst{6},
   M.~J.~Hardcastle\inst{7},
   M.~Iacobelli\inst{1},
   A.~P.~Mechev\inst{5},
   L.~K.~Morabito\inst{8},
   I.~Prandoni\inst{9},
   H.~J.~A.~R\"ottgering\inst{5},
   J.~Sabater\inst{4},
   T.~W.~Shimwell\inst{1},
   C.~Tasse\inst{10,11}
   \and
   W.~L.~Williams\inst{7}
   }
   \authorrunning{H. R. Stacey et al.}

   \institute{ASTRON, Netherlands Institute for Radio Astronomy, Oude Hoogeveensedijk 4, 7991 PD, Dwingeloo, The Netherlands
         \and
         Kapteyn Astronomical Institute, PO Box 800, 9700 AV Groningen, The Netherlands
         \and
Jodrell Bank Centre for Astrophysics, School of Physics and Astronomy, University of Manchester, Oxford Road, Manchester M13 9PL, UK
		\and
        SUPA, Institute for Astronomy, Royal Observatory, Blackford Hill, Edinburgh, EH9 3HJ, UK
        \and
Leiden Observatory, Leiden University, PO Box 9513, NL-2300 RA Leiden, the Netherlands
		\and
CSIRO Astronomy and Space Science, PO Box 1130, Bentley WA 6102, Australia
        \and
Centre for Astrophysics Research, University of Hertfordshire, College Lane, Hatfield AL10 9AB    
		\and
Astrophysics, University of Oxford, Denys Wilkinson Building, Keble Road, Oxford OX1 3RH, UK
        \and
INAF - Istituto di Radioastronomia, Via P. Gobetti 101, 40129 Bologna, Italy
		\and
GEPI, Observatoire de Paris, CNRS, Universit\'e Paris Diderot, 5 place Jules Janssen, 92190 Meudon, France 
		\and
Department of Physics \& Electronics, Rhodes University, PO Box 94, Grahamstown, 6140, South Africa
             }

   \date{Received July 29, 2018
   }

 
  \abstract
   {Determining the star-forming properties of radio-quiet quasars is important for understanding the co-evolution of star formation and black hole accretion. Here, we present the detection of the gravitationally-lensed radio-quiet quasars SDSS~J1055+4628, SDSS~J1313+5151 and SBS~1520+530 at 144~MHz that fall in the HETDEX Spring Field targeted in the LOFAR Two-metre Sky Survey (LoTSS) first full data release. We compare their radio and far-infrared luminosities relative to the radio--infrared correlation and find that their radio luminosities can be explained by star formation. The implied star formation rates derived from their radio and infrared luminosities are between 20 and 300~M$_{\odot}$~yr$^{-1}$. These detections represent the first study of gravitationally lensed sources with LOFAR, opening a new frequency window for investigating the star-forming properties of high redshift quasars at radio wavelengths. We consider the implications for future data releases and estimate that many of the objects in our parent sample will be detected during LoTSS, significantly increasing the fraction of gravitationally lensed radio-quiet quasars with radio detections.}
   

   \maketitle
%

\section{Introduction}

The star formation history of galaxies is thought to be strongly linked with black hole accretion, as evidenced through the peak in the activity of these two processes at $z\sim2$ \citep{Madau:2014}, and through the correlation between bulge luminosity and black hole mass \citep{Magorrian:1998,Ferrarese:2000,Gebhardt:2000}. Hydrodynamical simulations suggest that feedback from active galactic nuclei (AGN) is required to regulate star formation in the most massive galaxies in the Universe \citep{DiMatteo:2005,Springel:2005,Bower:2006}. The mechanism of AGN feedback may take the form of jets or winds, which mechanically eject gas from the host galaxy and deplete the reservoir of cold gas, or through radiative heating, which ionises the cold gas to quench star formation \citep[see][for review]{Fabian:2012}. A key test of this evolutionary scenario, therefore, is the star-forming properties of quasar host galaxies, most of which are expected to be dusty and gas-rich \citep{Hopkins:2005}.

Investigations of star formation in the high redshift Universe often focus on the far-infrared (FIR), which probes the cold dust of the interstellar medium (ISM) in the star-forming disk, or on radio wavelengths, which provide a view of star formation that is unobscured by dust. A major challenge in characterising star formation in quasar host galaxies has been sensitivity limitations. For the case of FIR--sub-mm instruments, these have historically detected the most extreme star-forming--quasar composites at high redshifts \cite[for example]{Isaak:2002,Priddey:2003,Pitchford:2016}. Similarly, radio synchrotron emission from all but the most extreme star-forming regions is faint and hard to detect. While progress in the FIR--sub-mm is being made with the Atacama Large (sub-)Millimetre Array (ALMA; e.g. \citealt{Stanley:2018}), and in the radio with deep, wide-field surveys such as with the Karl G. Jansky Very Large Array (VLA; e.g. \citealt{Smolcic:2017}) and the Low Frequency Array (LOFAR; \citealt{Rottgering:2006}), in both of these wavelength regimes there may also be a contribution due to black hole accretion, that is, from radio core/jet emission or AGN-heated dust emission, which is difficult to disentangle \citep[e.g.][]{Zakamska:2016}.

Understanding the relative contributions of star formation and black hole accretion is evidently important, not only to infer star formation rates (SFRs), but to understand the AGN feedback mechanisms involved. The well-established radio--infrared correlation for star-forming galaxies has been employed to determine if there is a radio excess due to radio jet emission \citep{Sopp:1991}. For radio-bright, jetted radio sources, we expect an excess of radio emission, however the majority of quasars are radio-`quiet' (i.e. have faint or undetected radio counterparts). Previous studies of optically-selected AGN have had conflicting results on their primary radio emission mechanism; some find this emission largely consistent with star formation \citep{Barthel:2006,deVries:2007,Condon:2013}, while others find that the AGN dominates \citep{White:2015,White:2017}. Studies of radio-selected samples also point towards primarily star formation \citep{Padovani:2011,Bonzini:2013,Bonzini:2015} or AGN emission \citep{Maini:2016,Herrera-Ruiz:2016,Herrera-Ruiz:2017} or a composite of both \citep{Delvecchio:2017}. These studies have typically focused on low redshift AGN or brighter-selected sources, or have detected fainter sources through stacking the optical positions of many undetected sources.

By taking advantage of the magnification effect of gravitational lensing, we can probe the FIR and radio properties of individual sources at high redshifts that otherwise would not be detectable. \cite{Stacey:2018} reported the results from a survey of 104 gravitationally-lensed quasars ($z\sim1$--4) that were observed with the SPIRE instrument aboard the {\it Herschel Space Observatory} at 250, 350 and 500~$\upmu$m, with which intrinsic flux densities below the confusion limit of the telescope could be probed. Under the assumption that the cold dust heating was due to star formation, they derived SFRs for the majority of the sample and upper-limits for the remaining sources. In general, high levels of dust-obscured star formation were found with a median of $120^{+160}_{-80}\ {\rm M_{\odot}\ yr^{-1}}$. While largely consistent with evolutionary models \citep[e.g.][]{Mancuso:2016}, the SFRs in the sample reach up to $10^4~{\rm M_{\odot}\ yr^{-1}}$ in the most extreme case. The radio-quiet quasars in the sample were found to be scattered close to the radio--infrared correlation, and \citeauthor{Stacey:2018} did not find a general infrared excess that would suggest there is a significant contribution to the FIR luminosity from AGN heating. In addition, the median cold dust temperature of the sample was $38_{-5}^{+12}$~K, which is consistent with the population of starburst galaxies at similar redshifts, and only a handful of the lensed quasars in the sample had dust temperatures larger than 50~K, which would suggest additional heating of the dust by the AGN. These high SFRs are difficult to reconcile with the expected luminosity functions of starburst galaxies \citep{Symeonidis:2018} and are in tension with current cosmological simulations \citep[e.g.][]{Dave:2010}. Thus, independent measurements of the star formation and AGN activity are needed.

However, such investigations are limited by the fact that the majority of the sample are currently undetected at radio wavelengths; almost all of the radio data for the radio-quiet sub-sample, as for most studies of large samples of quasars, come from the Faint Images of the Radio Sky at Twenty Centimeters (FIRST) survey \citep{Becker:1995} or the NRAO VLA Sky Survey (NVSS) \citep{Condon:1998}, which have detection limits of $S_{\rm 1.4~GHz} > 1$~mJy and $S_{\rm 1.4~GHz} > 2.5$~mJy, respectively. As optically-thin synchrotron emission from star-forming galaxies or AGN typically has a negative spectral index, observed radio luminosities are larger at lower frequencies. Therefore, using observations with the new low-frequency interferometric arrays, such as LOFAR \citep{vanHaarlem:2013}, we can increase our rate of detection at the same given sensitivity. As there are currently no survey instruments capable of observing at 40--100~$\upmu$m, where AGN-heated dust at $z\sim2$ dominates in the observed-frame, low-frequency radio observations present an alternative method of identifying these extreme cases where the radiative output from the AGN is heating the cold dust at FIR wavelengths.

In this paper, we report on the radio properties of three gravitationally lensed quasars from the \cite{Stacey:2018} sample that are detected with the LOFAR Two-metre Sky Survey (LoTSS, \citealt{Shimwell:2018}) of the HETDEX Spring Field. Our aim is to determine whether the radio emission from these quasars is consistent with what we would expect from the radio--infrared correlation, further adding to the case that these objects are also undergoing extreme bursts of star formation. Alternatively, we may find evidence for a radio or FIR excess, which would suggest that the AGN is contributing to the radio or heated dust emission, respectively. Either of these outcomes would also be interesting, as these quasars are thought to be radio-quiet, and detecting AGN-heated dust is difficult due to the thermal spectrum peaking at FIR wavelengths much shorter than the atmospheric cut-off. In Section~\ref{section:obs}, we review the quasars, and in Section~\ref{section:data}, we summarise the radio data. In Section~\ref{section:results}, we compare the radio and FIR properties relative to the radio--infrared correlation. The implications of our findings and future prospects are discussed in Section~\ref{section:discussion}. Finally, we present our conclusions in Section~\ref{section:conclusions}.

Throughout, we assume the \cite{Planck:2016} instance of a flat $\Lambda$CDM cosmology with $H_{0}= 67.8~$km\,s$^{-1}$ Mpc$^{-1}$, $\Omega_{\rm M}=0.31$ and $\Omega_{\Lambda}=0.69$.

\section{Sample}
\label{section:obs}

The targets of our pilot study are three lensed quasars from the {\it Herschel}/SPIRE parent sample of \cite{Stacey:2018} that fall in the LoTSS observations of the HETDEX Spring Field. All of these quasars were discovered at optical wavelengths, have two lensed images, and are undetected at 1.4~GHz in FIRST \citep{Becker:1995}. Dust temperatures and FIR-derived luminosities and SFRs (uncorrected for lensing magnification) are summarised in Table~\ref{table:Lfir}. The lensing magnifications of the radio and FIR emission for these systems are unknown, but are typically a factor of $\sim10$ based on analyses of high-resolution data of other lensed AGN (see \citealt{Stacey:2018} for details).

\subsection{SDSS~J1055+4628}

SDSS~J1055+4628 (SDSS~J105545.45+462839.4) was discovered in the Sloan Digital Sky Survey (SDSS) Quasar Lens Search (SQLS) by \cite{Kayo:2010}. The quasar is at $z_s=1.25$ and is gravitationally lensed by a foreground galaxy at $z_l=0.39$. The separation between the two lensed images is $1.15\arcsec$. This system was undetected in {\it Herschel}/SPIRE, implying a limit on the lensing-corrected SFR of $<44~{\rm M_{\odot}\ yr^{-1}}$, assuming the median fitted dust temperature of the parent sample and a magnification factor of 10.

\subsection{SDSS~J1313+5151}

SDSS~J1313+5151 (SDSS~J131339.98+515128.4) was discovered in the SQLS by \cite{Ofek:2007}. The background quasar is at $z_s=1.88$ and is gravitationally lensed into two images with a separation of $1.24\arcsec$ by a foreground galaxy at $z_l=0.19$. Heated dust emission that is assumed to be from the quasar host galaxy was detected only in the 250~$\upmu$m band, and is equivalent to a lensing-corrected SFR of $160^{+170}_{-80}~{\rm M_{\odot}\ yr^{-1}}$, based on the median and distribution of dust temperatures of the sample, and assuming a typical magnification of 10.

\subsection{SBS~1520+530}

SBS~1520+530 is a gravitationally lensed quasar at $z_s=1.86$ that was discovered in the Second Byurakan Survey \citep{Chavushyan:1997}. The primary lensing galaxy is at $z_l=0.72$ \citep{Burud:2002}. There is also evidence for a secondary lensing galaxy with a photometric redshift of $z_l\simeq0.9$ \citep{Faure:2002}. The two lensed images are separated by $1.59\arcsec$. \citet{Stacey:2018} detected dust emission, likely from the quasar host galaxy, at 250~$\upmu$m and 350~$\upmu$m, and blended emission at 500~$\upmu$m that included flux from nearby stars. The fitted dust temperature is $46.2^{+12.7}_{-7.2}$~K and the derived lensing-corrected SFR is $190^{+100}_{-60}~{\rm M_{\odot}\ yr^{-1}}$ assuming a typical magnification of 10. The peak of the dust emission is not well constrained, hence there are large uncertainties in the derived dust temperature and FIR luminosity for this object.

\section{Data}
\label{section:data}

\subsection{LOFAR LoTSS-DR1 data}

The LoTSS-DR1 data set is described in detail by \citet{Shimwell:2018}. To summarise, the first public data release of high quality survey images from LOFAR encompassed 424 square degrees that are coincident with the HETDEX Spring Field. In total, 325\,694 radio sources were detected at the 5$\sigma$-level with an angular resolution of $6\arcsec$. The median rms noise is $\sim71~\upmu$Jy~beam$^{-1}$, which is equivalent to a factor of about 10 times improvement in sensitivity when compared to the FIRST survey, accounting for the typical spectral index of synchrotron emission ($\alpha_{144}^{1400} \sim-0.7$) between 144~MHz and 1.4~GHz.

We first carried out a search of the LoTSS-DR1 sky coverage for known gravitationally lensed quasars, finding that the positions of the three targets described in Section~\ref{section:obs} were within the survey footprint. The properties of the radio-counterparts of these objects were then extracted from the LOFAR data using the software {\sc PyBDSF} \citep{Mohan:2015}, which fits 2-dimensional Gaussian models to the image-plane maps, as described by \cite{Shimwell:2018}. The fitted positions of these sources in the LOFAR data are within the errors of the optical positions given by SDSS for SDSS~J1055+4628 and SDSS~J1313+5151, and from observations with the {\it Hubble Space Telescope} ({\it HST}) for SBS~1520+530 \citep{Faure:2002}.

\begin{figure}
\includegraphics[width=0.435\textwidth]{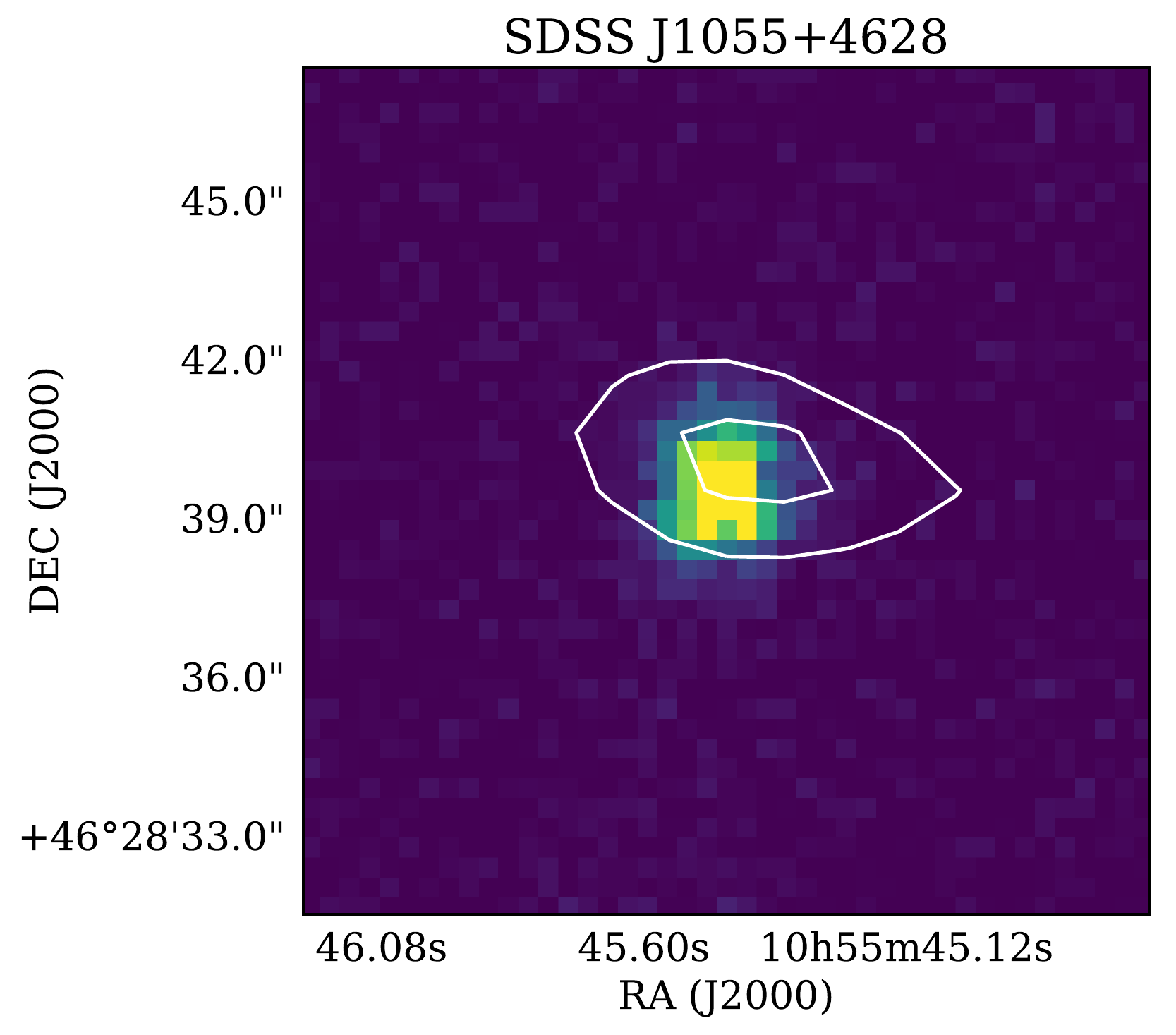}
\includegraphics[width=0.435\textwidth]{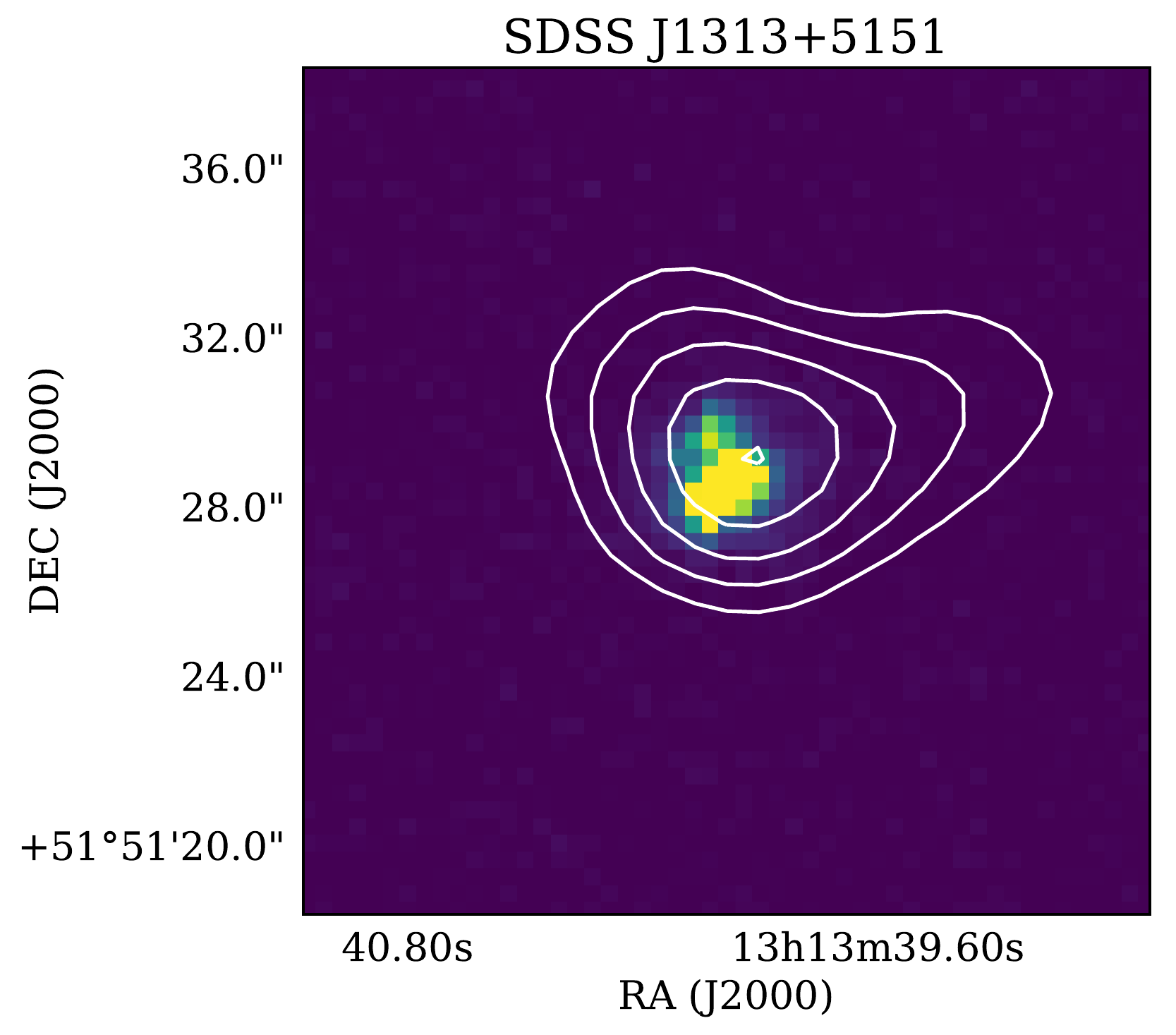}
\includegraphics[width=0.435\textwidth]{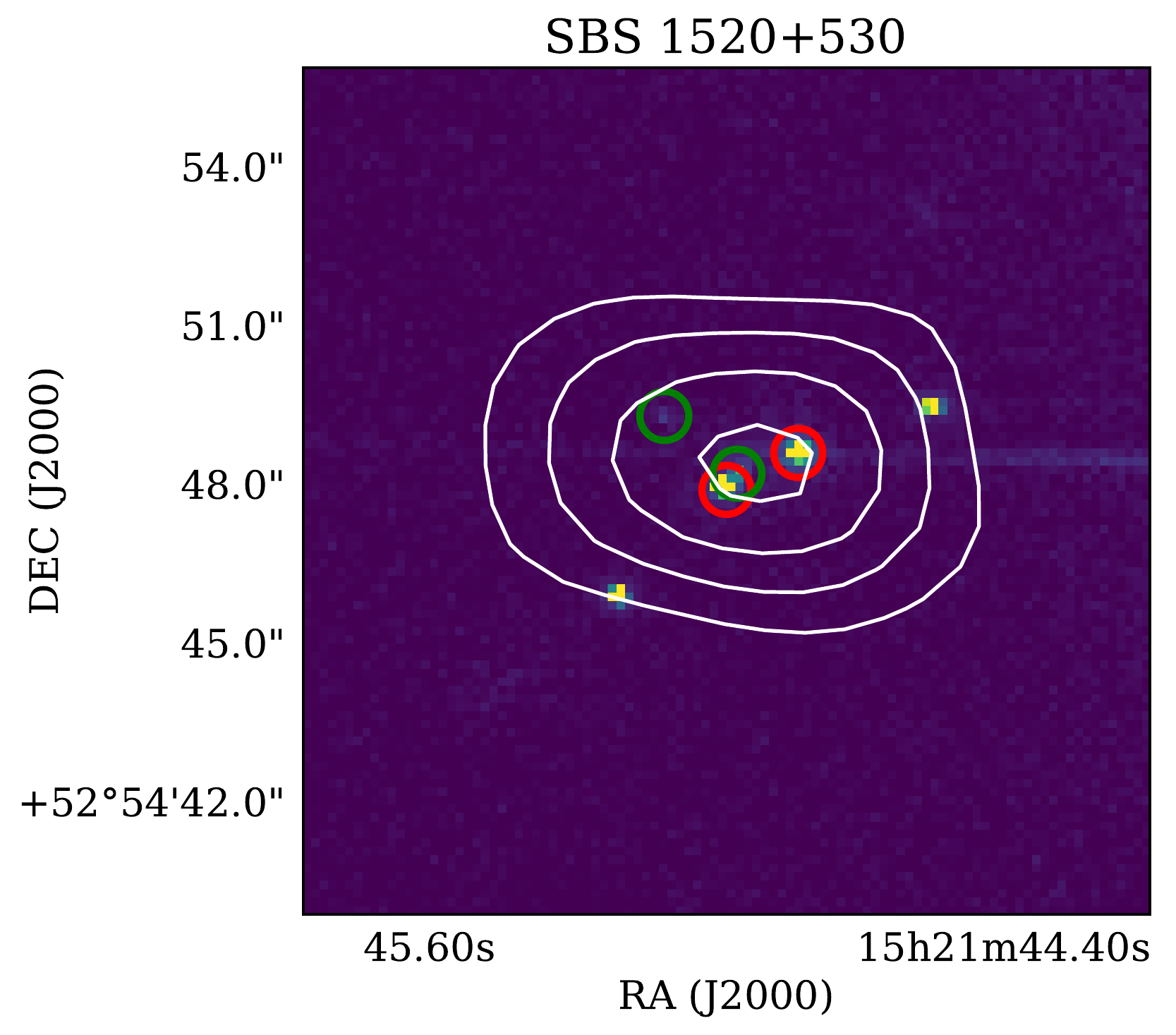}
\caption{The optical counterparts of the three gravitationally-lensed quasars, with the 144~MHz LoTSS contours overlaid. The contours shown are at the $3\sigma$, $4\sigma$, $5\sigma$, $6\sigma$ and $7\sigma$ level, where $\sigma$ is the off-source rms noise (given in Table~\ref{table:fits}). For SDSS~J1055+4826 and SDSS~J1313+5151, we show the SDSS $g$-band images. For SBS~1520+350, we show the {\it HST} $V$-band (F555W filter). For SBS~1520+350, the quasar images are indicated with the red circles, and the primary and secondary lensing galaxies are marked with the green circles; the unlabelled objects are foreground stars.}
\label{fig:images}
\end{figure}

The best quality optical imaging available for the three targets, with the contours from the LOFAR imaging overlaid, are presented in Figure~\ref{fig:images}, and the properties of the 144~MHz radio emission derived from the Gaussian fits are given in Table~\ref{table:fits}.

\subsection{Archival VLA data}

Previous investigations with the VLA have targeted SBS~1520+530; it was observed on 1997 August 28 at 1.4~GHz, with 20~minutes on source, under programme AH632. These data were extracted from the NRAO archive\footnote{https://archive.nrao.edu} and calibrated using the Common Astronomy Software Applications ({\sc casa}) package in a standard way, using 3C286 to set the absolute flux scale and nearby phase calibrators to trace and remove phase and amplitude variations. A 2$^\prime$ error in the pointing position of the phase calibrator was corrected for during this process. 

The rms noise level in the map is 65~$\upmu$Jy~beam$^{-1}$, although there are artefacts at the 200--300~$\upmu$Jy~beam$^{-1}$ level due to a combination of residual phase errors and incomplete {\it u-v} coverage. There is a positive spike of 280~$\upmu$Jy~beam$^{-1}$ around the position of the source, but due to the quality of the map it is not clear whether this is a noise artefact. Therefore, we assume 280~$\upmu$Jy as an upper limit for the flux density of SBS~1520+530 at 1.4 GHz.

In addition, only SBS~1520+530 is currently covered in the 3~GHz VLA Sky Survey (VLASS) preliminary data release \citep{Myers:2018}. The object is not detected with an rms noise level of 110~$\upmu$Jy~beam$^{-1}$.

\begin{table*}[ht]
\def\arraystretch{1.2}
\caption{The rest-frame dust temperature ($T_{\rm dust}$) and FIR luminosity ($\mu L_{\rm FIR}$), radio-infrared factor ($q_{\rm IR}$), IR-derived apparent SFR ($SFR_{\rm IR}$) for the three gravitationally-lensed quasars in this work. We also give the radio-derived apparent SFRs using the \cite{Calistro-Rivera:2017} radio--infrared correlation at 150~MHz ($SFR_{\rm 150}$) and the \cite{Ivison:2010} correlation at 1.4~GHz ($SFR_{\rm 1400}$). $T_{\rm dust}$, log~$\mu SFR_{\rm IR}$ and $\mu L_{\rm FIR}$ are as reported by \cite{Stacey:2018}. As the lensing magnification of the radio and FIR is not known for these objects, luminosities and SFRs given here are uncorrected for magnification ($\mu$).}
\begin{tabular*}{0.89\textwidth}{ p{3cm} | p{1.7cm} p{1.7cm} p{1.7cm} p{1.7cm} p{1.7cm} p{1.7cm}}
Source  & $T_{\rm dust}$ & log~$\mu L_{\rm FIR}$ & $q_{\rm IR}$ & log~$\mu SFR_{\rm IR}$      & log~$\mu SFR_{\rm 150}$ 
& log~$\mu SFR_{\rm 1400}$ \\
        & (K)            & ($\rm{L_{\odot}}$)    &              & ($\rm{M_{\odot}~yr^{-1}}$)  & ($\rm{M_{\odot}~yr^{-1}}$)  
& ($\rm{M_{\odot}~yr^{-1}}$)  \\
\hline
SDSS~J1055+4628 & - & $<12.1$ & $<2.3$ & $<2.6$ & $2.4\pm0.1$ & $2.7\pm0.1$\\
SDSS~J1313+5151 & - & $12.7^{+0.4}_{-0.2}$ & $2.5^{+0.5}_{-0.2}$ & $3.2^{+0.4}_{-0.2}$ & $2.8\pm0.1$ & $3.1\pm0.1$\\
SBS~1520+350 & $46^{+13}_{-7}$ & $12.8^{+0.2}_{-0.1}$ & $2.3^{+0.2}_{-0.2}$ & $3.3^{+0.2}_{-0.1}$ & $3.1\pm0.1$ & $3.4\pm0.1$
\label{table:Lfir}
\end{tabular*}
\end{table*}

\section{Results}
\label{section:results}

\subsection{LOFAR detections}

All three gravitationally-lensed quasars in the HETDEX footprint are detected in the LoTSS imaging between 6 to $9\sigma$, based on the total flux-density, with peak surface brightness at the 4 to $7\sigma$-level (see Figure~\ref{fig:images}). The quality of the data for SBS~1520+530 is lower than for the other two targets, as it is located at the edge of the field where the primary beam response is lower and is also affected by the residual side-lobes from a bright source nearby. While the emission from all three gravitational lens systems appears to be marginally resolved, unresolved sources are artificially broadened in these data, perhaps due to a combination of ionospheric effects, astrometric errors or calibration quality, as discussed by \cite{Shimwell:2018}. Criteria established by \cite{Shimwell:2018} classify these sources as unresolved, which is consistent with the $6\arcsec$ beam-size of the observations and the $1.2\arcsec$ to $1.6\arcsec$ separations of the lensed images. We use the integrated flux density for the rest of our analysis, as the peak surface brightness of a point source is likely to be underestimated.

With the 144~MHz flux densities provided by LoTSS, we are now able to estimate the rest-frame 1.4~GHz luminosities for these three sources. We use a simple power-law with spectral index $\alpha$,
\begin{equation}
S_{\nu} \propto \nu^{\alpha},
\label{eq:spectrum}
\end{equation}
to describe the flux density ($S_\nu$) as a function of frequency ($\nu$) for optically-thin synchrotron emission. We assume a spectral index between 144~MHz and 1.4~GHz of $\alpha^{1400}_{144}=-0.7$ to extrapolate to the observed-frame 1.4~GHz flux density. This assumption is consistent with the typical 150~MHz to 1.4~GHz spectral index found for both AGN and star-forming galaxies by \cite{Calistro-Rivera:2017} and \cite{Gurkan:2018b}. It is also consistent, in the case of SBS~1520+530, with the archival data from the VLA at GHz frequencies: the possible detection at 280~$\mu$Jy derived from the 1.4~GHz VLA map would suggest a spectral index of $\alpha_{144}^{1440} = -0.78$ for this object. This is also consistent with the non-detection in VLASS, which implies a limit on the spectral index of $\alpha_{144}^{3000}<-0.6$ assuming an object could be detected at the 3$\sigma$-level.

Finally, we determine the rest-frame 1.4 GHz luminosity using,
\begin{equation}
L_{\rm 1.4~GHz, rest} = 4 \pi \, D_L^2 \, S_{\rm 1.4~GHz, obs} \, (1+z_s)^{-(1 +\alpha)},
\end{equation}
where $D_L$ is the luminosity distance.

We also check for any uncertainties in the absolute flux scaling of LoTSS by comparing the relative flux ratios between common sources in our target fields and the TIFR GMRT Sky Survey (TGSS; \citealt{Intema:2017}) catalogue (see \citealt{Shimwell:2018} for details). We find that, for the fields of SDSS~J1055+4628 and SBS~1520+350, the LoTSS and TGSS flux-densities of bright objects agree to within a few percent. But for the field of SDSS~J1313+5151, the median difference may be as much as 20\%. This will have a limited impact on our analysis as the uncertainty in the radio--SFR relation is typically a factor of 2.

\begin{table*}[ht]
\def\arraystretch{1.2}
\caption{Parameters of the 2-dimensional Gaussian models fitted to the 144~MHz LoTSS image-plane data using {\sc PyBDSF}. We give the peak surface brightness ($I_{\rm peak}$) and the integrated flux density ($S_{\rm int}$), the size of the fitted Gaussians (not de-convolved from the LOFAR beam-size; $6\arcsec \times 6\arcsec$), and the off-source rms noise in the region of the targets. We also show the ratio of the flux-densities from LoTSS and TGSS for nearby radio sources in each field ($S_{\rm LoTSS}$/$S_{\rm TGSS}$).}
\centering
\begin{tabular*}{\textwidth}{ p{3cm} | p{1.7cm} p{1.7cm} p{1.7cm} p{1.7cm} p{1.7cm} p{1.7cm} p{1.7cm}}
Source 	& $I_{\rm peak}$ 	& $S_{\rm int}$ & Major~axis	& Minor~axis 	& PA 	& rms~noise & $S_{\rm LoTSS}$/$S_{\rm TGSS}$ \\ 
		& (mJy~beam$^{-1}$)	&(mJy)		& (arcsec)		& (arcsec)		& (deg.~E~of~N)	& (mJy~beam$^{-1}$) & \\
\hline
SDSS~J1055+4628 & $0.40\pm0.08$ & $0.76\pm0.12$ & $10.1\pm2.3$ & $6.7\pm1.2$ & $55\pm25$ & 0.07 & $1.02\pm0.24$ \\
SDSS~J1313+5151 & $0.41\pm0.06$ & $0.81\pm0.09$ & $9.4\pm1.5$ & $7.6\pm1.0$ & $79\pm32$ & 0.07 & $0.80\pm0.18$ \\
SBS~1520+350 & $1.07\pm0.17$ & $1.64\pm0.26$ & $9.2\pm1.7$ & $6.0\pm0.8$ & $84\pm21$ & 0.16 & $0.97\pm0.26$ 
\label{table:fits}
\end{tabular*}
\end{table*}

\begin{figure}
\includegraphics[width=0.5\textwidth]{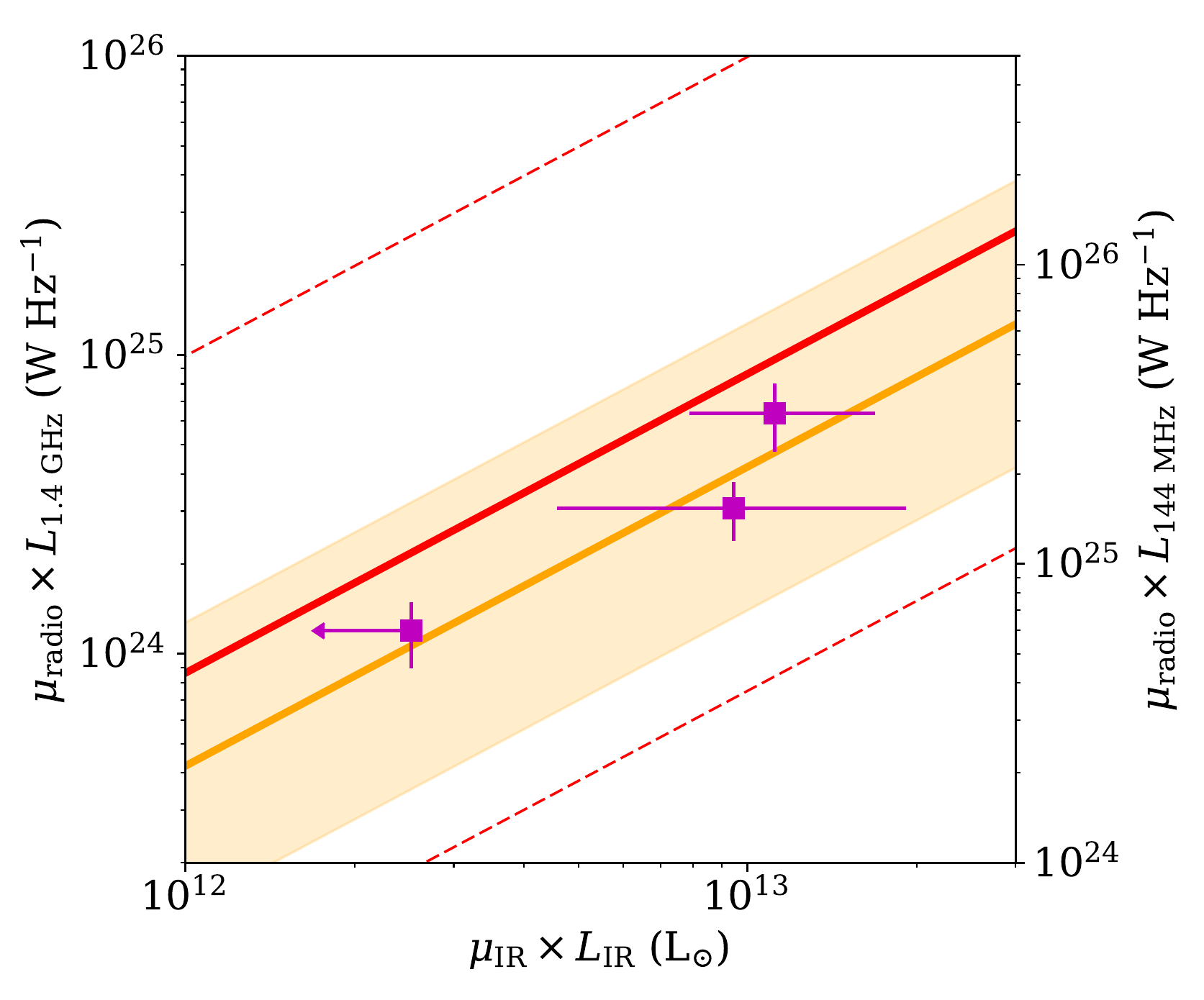}
\caption{The rest-frame infrared and 1.4~GHz luminosities of the three lensed quasars detected with LOFAR, uncorrected for the lensing magnification. The corresponding rest-frame 144~MHz luminosities are shown on the opposing axis, scaled by spectral index $\alpha^{1400}_{144}=-0.7$. The median $q_{\rm IR}$ for star-forming galaxies from \protect\cite{Ivison:2010} is shown in yellow; the shaded region is $2 \times \sigma_{\rm q_{IR}}$, where $\sigma_{\rm q_{IR}}$ is the intrinsic scatter. The solid red line shows the radio--infrared correlation at 150~MHz from \cite{Calistro-Rivera:2017} at $z\sim1.5$ scaled to 1.4~GHz, with the dashed line showing $2 \times \sigma_{\rm q_{IR}}$.}
\label{fig:FIRRC1GHz_lofar}
\end{figure}

\begin{figure*}
\centering
\includegraphics[width=0.75\textwidth]{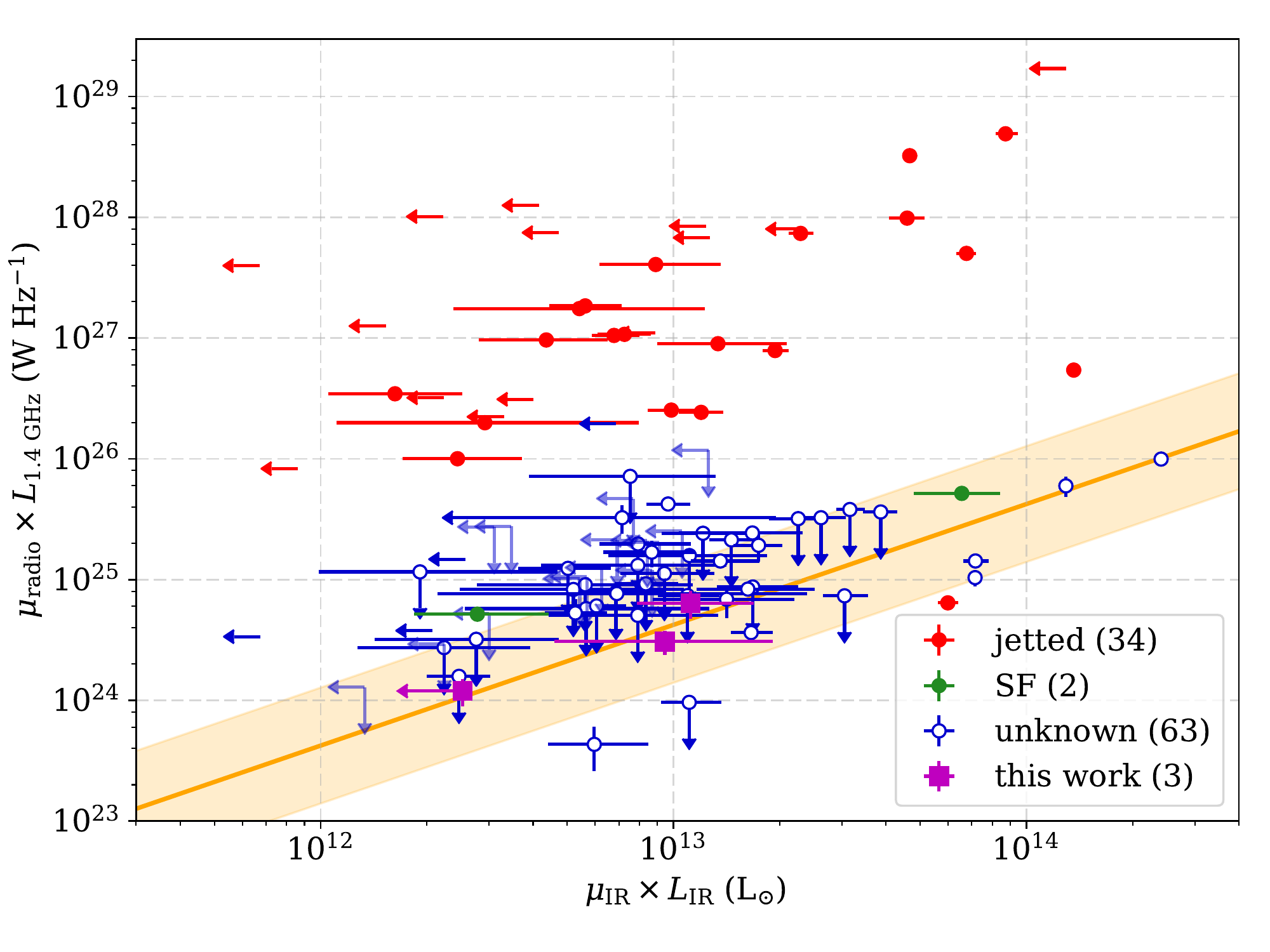}
\includegraphics[width=0.75\textwidth]{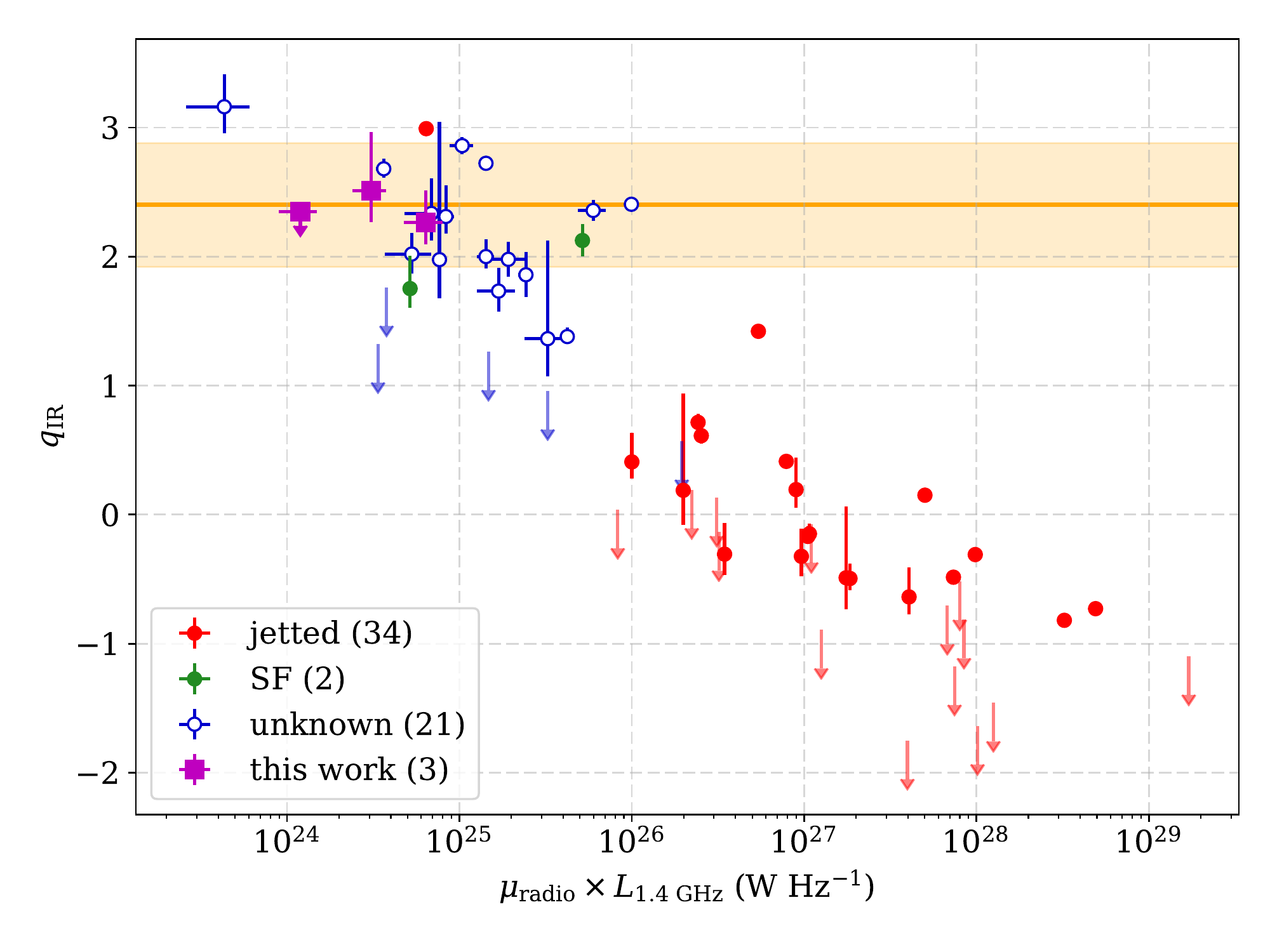}
\caption{The top panel shows the rest-frame infrared and extrapolated 1.4~GHz luminosities of the parent sample from \cite{Stacey:2018}, uncorrected for the lensing magnification. The bottom panel shows the $q_{\rm IR}$ values and extrapolated 1.4~GHz luminosities of those lensed quasars in the parent sample with a known radio detection. Quasars with jet-dominated radio emission are shown in red; quasars with radio emission known to be star-formation-dominated are shown in green; those with an undetermined radio emission mechanism are shown in blue. The three objects in this work are shown in purple. The median $q_{\rm IR}$ for star-forming galaxies from \protect\cite{Ivison:2010} is shown in yellow; the shaded region is $2 \times \sigma_{\rm q_{IR}}$.}
\label{fig:FIRRC1GHz}
\end{figure*}

\subsection{Radio--infrared correlation at 1.4~GHz}

The radio--infrared correlation is described by the parameter $q_{\rm IR}$, the ratio between the total infrared luminosity (8 to 1000~$\upmu$m; rest-frame) and the 1.4 GHz rest-frame luminosity, defined by \cite{Condon:1991} as,
\begin{equation}
q_{\rm IR} = \log_{10}\left (\frac{L_{\rm IR}}{3.75\times10^{12}\ L_{\rm 1.4\ GHz}}  \right ).
\label{eq:qir}
\end{equation} 
This relation falls naturally from the dust and radio emission being associated with star formation that is produced by the same stellar population \citep{Schober:2017}. We extrapolate from $L_{\rm FIR}$ (40 to 120~$\upmu$m; rest-frame) given by \cite{Stacey:2018} to the total infrared luminosity (8 to 1000~$\upmu$m; rest-frame) using the colour correction factor of 1.91 of \cite{Dale:2001} (i.e. $L_{\rm IR}=1.91\,L_{\rm FIR}$), calibrated for star-forming galaxies, to allow for the contribution from mid-infrared spectral features. Ideally, we would use a spectral energy distributions (SED) model or template which would also account for an AGN component (e.g. \citealt{Siebenmorgen:2015,Calistro-Rivera:2016}). However, our targets have no data beyond 22~$\upmu$m (e.g. from {\it Spitzer}/MIPS, {\it Herschel}/PACS) that would enable us to constrain contributions from the AGN torus emission to the FIR. We adopt the \citeauthor{Dale:2001} colour correction under the assumption that the FIR luminosity is associated with star formation, without a significant contribution from black hole accretion (an assumption that we will test here by comparison with the radio properties). For consistency with other analyses, we use $q_{\rm IR}=2.40 \pm 0.24$ \citep{Ivison:2010}, which was derived from 1.4~GHz radio and {\it Herschel}/SPIRE observations for luminous infrared galaxies at a similar redshift to our targets, and assume a \cite{Salpeter:1955} initial mass function (IMF) for conversion from luminosity to SFR.

We also consider the evolution of the radio--infrared correlation with redshift, for which there are conflicting reports in the literature. \cite{Ivison:2010} find a suggestion of evolution with redshift, which may be due to selection bias. \cite{Delhaize:2017} and \cite{Magnelli:2015} find statistically-significant evidence of evolution with redshift, however for the redshifts of our targets ($z\sim1.5$) the inferred $q_{\rm IR}\simeq2.3$--2.4 are consistent with \cite{Ivison:2010}.

As there is no high-resolution imaging for our targets, we do not have information on the lensing magnification of the radio ($\mu_{\rm radio}$) or dust emission ($\mu_{\rm IR}$). If the radio and FIR emission both originate from the star-forming disk and are co-spatial, their total magnifications will be similar. In this case, the ratio of the observed luminosities (and, hence, $q_{\rm IR}$) will be the same as the lensing-corrected (intrinsic) ratio. However, the relative magnification of any radio emission associated with the AGN will depend on the morphology and location of this relative to the FIR emission and the lensing caustics: while the radio emission in radio-quiet quasars associated with accretion is assumed to be compact, these can also exhibit diffuse radio structures (e.g. \citealt{Harrison:2015,Alexandroff:2016,Baldi:2018}). Here, we assume that the star formation is not preferentially magnified relative to any AGN emission. While high-resolution data will be needed to fully account for such a scenario through lens modelling, simulations by \cite{Serjeant:2012} suggest that the effect of differential magnification is not likely to be significant enough to alter our interpretation.

The rest-frame radio and infrared luminosities of our targets, relative to the expectations from the radio--infrared correlation, are shown in Figure~\ref{fig:FIRRC1GHz_lofar}. We find that for the gravitationally-lensed quasars detected with LoTSS, their infrared luminosities are consistent with the radio--infrared correlation derived at 1.4~GHz by \cite{Ivison:2010} and at 150~MHz by \cite{Calistro-Rivera:2017}. This suggests that the radio emission is mostly or entirely driven by star formation.

In Figure~\ref{fig:FIRRC1GHz}, we show the same data for the targets studied here, but also include the other 99 lensed quasars in the {\it Herschel}/SPIRE sample (the remaining 2 objects of the {\it Herschel}/SPIRE sample have no radio measurements; see \citealt{Stacey:2018}). Also in Figure~\ref{fig:FIRRC1GHz}, the $q_{\rm IR}$ values of the radio-detected quasars relative to their radio luminosities are shown. We have  defined sub-samples of the sources to indicate those that are jetted (34 objects that show evidence for radio-jets in high-angular-resolution imaging) and those that are non-jetted (2 objects with confirmed diffuse radio emission that is expected to be due to star formation). Due to the lack of very long baseline interferometric imaging for the whole sample, the nature of the radio emission from most of the quasars, including the three sources detected by LoTSS, is unconfirmed.

\subsection{Radio-derived star formation rates}

Under the assumption that the radio continuum is attributed to only star formation, the radio luminosity can be used to estimate the SFR by inferring the IR luminosity from the radio--infrared correlation. For consistency with \cite{Stacey:2018}, we derive SFRs using the methodology of \cite{Kennicutt:1998}, assuming a Salpeter initial mass function,
\begin{equation}
SFR~{\rm (M_{\odot}~yr^{-1})} = \frac{L_{\rm IR}}{5.8\times10^{9}} ,
\label{eq:kennicutt}
\end{equation}
where $L_{\rm IR}$ is in units of L$_{\odot}$.

From the radio--infrared correlation of \citeauthor{Ivison:2010}, we expect SFRs (uncorrected for lensing magnification) of $500\pm130$, $1300\pm290$ and $2700\pm700$~${\rm M_{\odot}\ yr^{-1}}$ for SDSS~J1055+4628, SDSS~J1313+5151 and SBS~1520+530, respectively. In the case of SDSS~J1055+4628, this is the first estimate of the SFR for this radio-quiet quasar.

We also derive SFRs from the 144~MHz continuum using the radio--infrared correlation at $z\sim1.5$ from \cite{Calistro-Rivera:2017}, for which we also use the relation in Equation~\ref{eq:kennicutt} for consistency with 1.4~GHz. In this case we find apparent SFRs of $250\pm60$, $650\pm140$ and $1300\pm340$~${\rm M_{\odot}\ yr^{-1}}$ for SDSS~J1055+4628, SDSS~J1313+5151 and SBS~1520+530, respectively. These are also within the errors of the SFRs we derive from the FIR luminosity for SDSS~J1313+5151 and SBS~1520+530, however a they are factor of 2 lower than the SFR derived from the \citeauthor{Ivison:2010} correlation at 1.4~GHz. This is perhaps unsurprising given the large scatter of the correlation at 150~MHz, but could relate to thermal contributions in either band. \cite{Gurkan:2018a} find a significant difference in the radio-derived SFRs between 1.4~GHz and 150~MHz, which they attribute to a contribution from thermal (free-free) emission at GHz frequencies. However, it is not clear that the 150~MHz luminosities are free from thermal processes either. Indeed, \cite{Calistro-Rivera:2017} found evidence for flattening spectra at lower radio frequencies, a feature that has been also observed in low-redshift star-forming galaxies (e.g. \citealt{Marvil:2015,Galvin:2018}). Such a scenario could also account for differences in the radio-derived SFRs calibrated at low and high frequencies.

Given the redshifts of our targets, the 144~MHz LoTSS data from LOFAR are probing the rest-frame $\sim400$~MHz emission, where the suppression due to free-free absorption is expected to be less, and any systematic error due to extrapolating the radio spectrum to 1.4~GHz should be limited. Therefore, we use the SFRs derived from the 1.4~GHz luminosity for the remainder of our analysis.

\begin{figure*}
\centering
\includegraphics[width=0.9\textwidth]{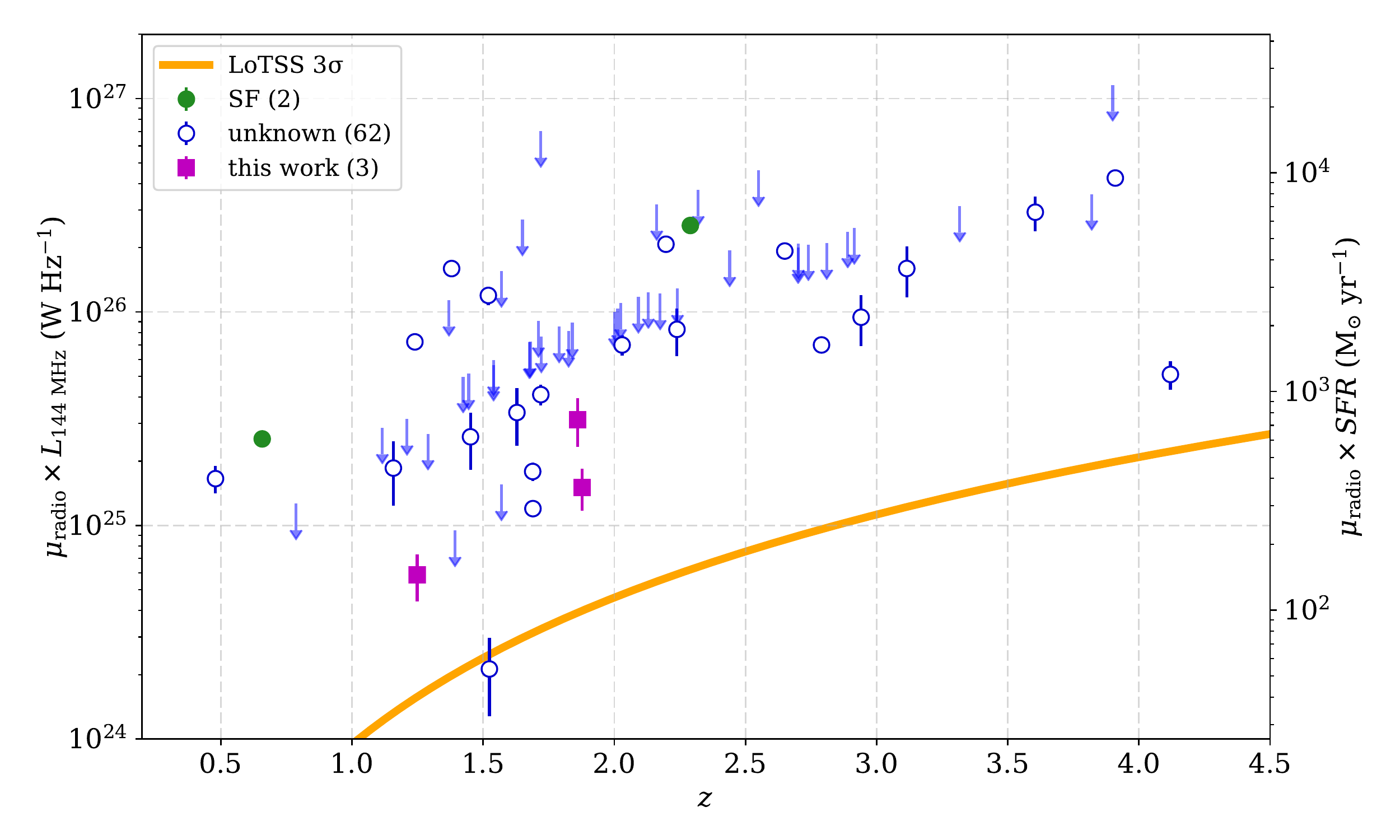}
\caption{The extrapolated rest-frame 144~MHz luminosities of 67 quasars in the parent sample with known redshifts and radio measurements that have not been confirmed as jetted from high angular-resolution radio imaging at mas-scales. Quasars with star-formation-dominated radio emission are shown in green; those with an undetermined radio emission mechanism are shown in blue. The quasars in this work are shown in purple. The orange line shows a $3\sigma$ surface brightness threshold in luminosity for an unresolved point source detected in LoTSS, assuming a median rms noise of 0.07~mJy~beam$^{-1}$. The SFR implied from the 1.4~GHz luminosity is shown on the right-hand y-axis using the radio--infrared correlation of \cite{Ivison:2010}. Most of the upper limits are from FIRST at the $5\sigma$-level, thus the LoTSS threshold shows a clear advantage in sensitivity.}
\label{fig:lofar-lum}
\end{figure*}

\section{Discussion}
\label{section:discussion}

\subsection{Radio emission mechanism of radio-quiet quasars}

For the two lensed quasars detected with {\it Herschel}/SPIRE, we find that these objects lie on the radio--infrared correlation, therefore the radio emission at 144~MHz can be explained by star formation. This implies that there is no significant AGN-heating of the cold dust, under the assumption that the radio emission is also dominated by star formation. While we cannot determine this for the lensed quasar that was not detected with {\it Herschel}/SPIRE, which could still have a radio excess, we do not find evidence of a FIR excess here. The SFRs implied from their projected radio luminosities are consistent within the errors and upper limit of the SFRs that were derived from the FIR emission, and they are in the range 20 to 300~M$_{\odot}$~yr$^{-1}$, assuming a typical lensing magnification of 10.

A study of unlensed quasars in the LoTSS first data release by \cite{Gurkan:2018b} find that optically-selected quasars at $z<1.0$, with a comparable radio luminosity to our sample, are scattered close to the radio--infrared correlation, consistent with our findings. They do not detect any distinct bi-modality in the radio properties of the quasar population, the lack of which may imply that this population has a composite of emission related to both black hole accretion and star formation, where star formation begins to dominate at low radio luminosities. This scenario is consistent with other studies focusing on deep observations of single fields, such as \cite{Delvecchio:2017}, who find that star formation accounts for most of the radio emission in the majority of AGN based on a large sample in the VLA-COSMOS 3 GHz survey. These findings may also be reconciled with studies such as by \cite{White:2017}, where half of the sample of 70 radio-quiet quasars are detected, which in general were found to have a radio excess. The differing conclusions may relate to how the studies are designed; if there is composite emission in most quasars, deep observations with very long baseline interferometry at cm-wavelengths will detect the compact emission associated with low-luminosity AGN activity in many cases \cite[e.g.][]{Herrera-Ruiz:2017,Radcliffe:2018}, and observations to test this for our sample of lensed quasars are under way.

High levels of star formation in quasar host galaxies are expected according to the paradigm of galaxy evolution in which star formation and AGN co-evolve. A lack of radio activity related to black hole accretion has implications on the AGN feedback mechanisms by which these quasars interact with their host galaxies; in the small sample studied here it seems that jets do not play a dominant role. However, radiative feedback from the AGN may have an effect, which is not traced by these data. We do note that for the sample of quasars studied here, we find both on-going AGN activity in the form of optical emission and evidence for on-going star formation, based on the radio and FIR emission. This suggests that star formation has not been halted in these objects. Additional radio imaging for the larger sample of lensed quasars studied by \citet{Stacey:2018}, which includes those objects with predicted SFRs $>10^3$~M$_{\odot}$~yr$^{-1}$ based on their FIR emission, will give a more complete view on the level of star formation activity and AGN heating of the dust within quasar host galaxies. In this way, we can confirm or discount the extreme rates of star formation inferred from the FIR. In general, such studies would be complementary to those of field quasars, such as by \cite{Gurkan:2018b}, and can provide a more complete picture of the evolutionary role of AGN within galaxy formation.

\subsection{Implications for future radio surveys}

Recently, significant progress has been made in characterising the faint radio population with deep surveys of single fields, such as the VLA-COSMOS 3~GHz survey \citep{Smolcic:2017}. However, most lensed quasars still lie below the detection limits of all-sky surveys such as FIRST, and even forthcoming surveys such as VLASS. The steep negative spectrum of synchrotron radiation provides an efficient way to detect the faint emission from star formation at high redshifts with LOFAR. The detection in this work of all three lensed quasars in the HETDEX field implies that future LoTSS data releases will enable a large sample of radio-quiet lensed quasars to be detected. If these sources do not have a significant excess from AGN-heated dust, we can expect these lensed quasars to be consistent with star-formation-dominated radio emission in the absence of radio jet emission. If we do not detect these sources, we can conclude that the AGN are heating dust on large-scales in their host galaxies. Further studies of the cold dust at high angular resolution with ALMA and the NOrthern Extended Millimetre Array (NOEMA) would help us to better understand the effect AGN have on the ISM of galaxies. The LoTSS survey being carried out with LOFAR could provide a novel method for selecting such targets.

We extrapolate our 144~MHz detections to $\sim$GHz frequencies, in the same way as described in Section~\ref{section:results}, and find that the three quasars in our sample have implied flux densities at observed-frame 1.4~GHz of 0.1--0.2~mJy. This implies that we would require sensitivities 5--10 times better than FIRST to detect these sources. In Figure~\ref{fig:lofar-lum}, we show the extrapolated 144~MHz radio luminosities (including current upper limits) of the {\it Herschel}/SPIRE sample, including the three detections reported here, and a $3\sigma$ surface brightness threshold given the typical rms noise of LoTSS. We also show the inferred SFR from the 1.4~GHz luminosity (uncorrected for the lensing magnification). Where there are no detections or detections in only one radio band, we assume a spectral index of $\alpha_{144}^{1400} = -0.7$, as before, otherwise we extrapolate or interpolate from existing radio measurements to the rest-frame frequencies. Only one quasar that was previously detected at GHz frequencies from targeted observations is projected to be a non-detection at the $3\sigma$-level in LoTSS images. The significant detection of the three lensed quasars in this work suggests that many of the lensed quasars currently undetected in FIRST could be detected in future LoTSS data releases. The typical magnifications of the star formation in these systems is a factor of 10, suggesting that we will probe SFRs an order of magnitude lower than accessible in large surveys of unlensed systems at similar redshifts in LoTSS \citep{Gurkan:2018a,Gurkan:2018b}.

Finally, we note that without the capability to resolve the gravitational lens systems, we cannot be sure that there is no contribution to the measured radio emission from the foreground lensing galaxy. This has been found to be the case for $\sim10 \%$ of gravitational lens systems with resolved radio emission at cm-wavelengths  \citep{McKean:2005,McKean:2007,Wucknitz:2008,Jackson:2015}. With the forthcoming long baseline capabilities of the International LOFAR Telescope as part of LoTSS, we will be able to achieve $\sim0.2\arcsec$ angular-resolution and potentiality identify these relative contributions. Alternatively, high resolution, high sensitivity observations at cm-wavelengths could also detect if there is any radio emission from the lensing galaxies \citep[e.g.][]{Jackson:2015}.

\section{Conclusions}
\label{section:conclusions}

We have detected three gravitationally-lensed radio-quiet quasars in the HETDEX Spring Field with LOFAR at 144~MHz that were undetected at $\sim$GHz radio frequencies. These quasars were previously observed with {\it Herschel}/SPIRE, with which SFRs could be derived for two and an upper limit placed on the third, based on their FIR SEDs. In this paper, we derived rest-frame luminosities from our radio measurements and, by comparing with the radio--infrared correlation, find that the radio luminosities can be produced by star formation. The SFRs inferred from their projected 1.4~GHz luminosities are consistent with the rates derived from their total infrared luminosities. Overall, the radio luminosities, infrared luminosities, low dust temperatures and steep dust emissivities are all consistent with extreme levels of on-going star formation within our parent sample of lensed quasars. However, we find a factor of 2 lower SFRs when using luminosity relations calibrated at 150~MHz for low redshift star-forming galaxies relative to the relation calibrated at 1.4~GHz. The reasons for this are not clear, but may be due to thermal contributions in one or both radio bands, which would have the effect of increasing the flux measurements due to free-free emission at 1.4~GHz and decreasing them due to free-free absorption at 150~MHz.

The nature of the radio emission from radio-quiet quasars has been the subject of controversy, with studies yielding different conclusions on the primary emission mechanism and whether the physical difference between the radio-loud and radio-quiet quasars reflects a true dichotomy in the quasar population. Our results point towards star-formation-dominated radio emission for the sources in our pilot sample, which is consistent with a similar study of unlensed radio-quiet quasars at low redshifts in this LoTSS data release by \cite{Gurkan:2018b}. A larger sample will be required to understand how robust these conclusions are for the quasar population in general. By extrapolating the 1.4~GHz luminosities of the parent sample to 144~MHz, we predict that many of the lensed quasars currently undetected at 1.4~GHz will be detected in future LoTSS data releases. As most of the radio-quiet lensed quasars in our parent sample are currently undetected, we expect to significantly increase the fraction with radio detections. Due to the magnification effect of gravitational lensing, we will be able to probe lower luminosities than otherwise accessible at $z\sim2$ and create a more complete picture of the radio-quiet quasar population at this epoch, when both the star formation and AGN activity peaked.

\begin{acknowledgements}
PNB and JS are grateful for support from the UK STFC via grant ST/M001229/1. KJD acknowledges support from the ERC Advanced Investigator programme NewClusters 321271. GG acknowledges the CSIRO OCE Postdoctoral Fellowship. MJH and WLW acknowledge support from the UK Science and Technology Facilities Council (STFC) [ST/M001008/1]. APM would like to acknowledge the support from the NWO/DOME/IBM programme ``Big Bang Big Data: Innovating ICT as a Driver For Astronomy'', project \#628.002.001. LKM acknowledges support from Oxford Hintze Centre for Astrophysical Surveys which is funded through generous support from the Hintze Family Charitable Foundation. This publication arises from research partly funded by the John Fell Oxford University Press (OUP) Research Fund. IP acknowledges support from INAF under PRIN SKA/CTA `FORECaST’. The LOFAR group in Leiden is supported by the ERC Advanced Investigator programme New-Clusters 321271. 
LOFAR, the Low Frequency Array designed and constructed by ASTRON, has facilities in several countries, that are owned by various parties (each with their own funding sources), and that are collectively operated by the International LOFAR Telescope (ILT) foundation under a joint scientific policy. The ILT resources have benefited from the following recent major funding sources: CNRS-INSU, Observatoire de Paris and Universit\'e d'Orl\'eans, France; BMBF, MIWF-NRW, MPG, Germany; Department of Business, Enterprise and Innovation (DBEI), Ireland; NWO, The Netherlands; The Science and Technology Facilities Council (STFC), UK. Part of this work was carried out on the Dutch national e-infrastructure with the support of the SURF Cooperative through grant e-infra 160022 \& 160152. The LOFAR software and dedicated reduction packages on \url{https://github.com/apmechev/GRID_LRT} were deployed on the e-infrastructure by the LOFAR e-infragroup, consisting of J. B. R. Oonk (ASTRON \& Leiden Observatory), A. P. Mechev (Leiden Observatory) and T. Shimwell (ASTRON) with support from N. Danezi (SURFsara) and C. Schrijvers (SURFsara). This research has made use of the University of Hertfordshire high-performance computing facility (\url{http://uhhpc.herts.ac.uk/}) and the LOFAR-UK computing facility located at the University of Hertfordshire and supported by STFC [ST/P000096/1].
The National Radio Astronomy Observatory is a facility of the National Science Foundation operated under cooperative agreement by Associated Universities, Inc.
This research made use of Astropy, a community-developed core Python package for Astronomy \citep{Astropy:2018}.
\end{acknowledgements}

\bibliographystyle{aa}
\bibliography{references}

\end{document}